\documentstyle[prd,aps,eqsecnum]{revtex}

\begin{document}
\input epsf
\draft
\twocolumn[\hsize\textwidth\columnwidth\hsize\csname
@twocolumnfalse\endcsname
\preprint{OU-TAP 91,SU-ITP-99-03, astro-ph/9901135, January 12, 1999}
\title{CMB in Open Inflation}
\author{Andrei Linde,$^1$ ~ Misao Sasaki,$^2$ and Takahiro Tanaka$^2$ }
\address{${}^1$Department of Physics, Stanford University, Stanford, CA
94305, USA}
\address{${}^2$Department of Earth and Space Science,
Graduate School of Science,}
\address{Osaka University, Toyonaka 560-0043, Japan}
\date{January 12, 1999}
\maketitle
\begin{abstract}
The possibility to have an infinite open inflationary universe inside a
bubble of a finite size is one of the most interesting realizations
extensively discussed in the literature. The original idea was based on
the theory of tunneling and bubble formation in the theories of a single
scalar field. However, for a long time we did not have any consistent
models of this type, so it was impossible to compare predictions of such
models with the observational data on the CMB anisotropy. The first
semi-realistic model of this type was proposed only very recently, in
hep-ph/9807493. Here we present the results of  our investigation of
the scalar and tensor perturbation spectra and the resulting CMB
anisotropy in such models. In all models which we have studied
there are no supercurvature perturbations.
The spectrum of scalar CMB anisotropies has a minimum at small\, $\ell $ and a  plateau at $\ell = O(10)$
for low $\Omega_0$.  Meanwhile tensor CMB anisotropies are peaked at
$\ell=2$. Relative magnitude of the scalar CMB spectra versus
tensor CMB spectra at small $\ell$ depends on the parameters of the
models.
\end{abstract}
\pacs{PACS: 98.80.Cq  \hskip 2.3cm OU-TAP 91~~~~~SU-ITP-99-03
\hskip 2.3cm  astro-ph/9901135}
\vskip2pc]

\section {Introduction}

Inflationary theory has a robust prediction: Our universe must be almost
exactly flat, $\Omega_0= \Omega_{\rm matter} + \Omega_\Lambda = 1 \pm
O(10^{-4})$. If this result is confirmed by observational data, we will
have a decisive confirmation of inflationary cosmology. However, what if
observational data show that the universe is open?

Until very recently, we did not have any consistent cosmological models,
inflationary or not, describing a homogeneous open universe. An
assumption that all parts of an infinite universe can be created
simultaneously and have the same value of energy density everywhere did
not have any justification. This problem was solved only after the
invention of inflationary cosmology.
 It was found that each bubble of a new phase formed during the false
 vacuum decay in inflationary universe   looks from inside like an
 infinite open universe \cite{CL,Gott}. The process of bubble formation
 in the false vacuum is described by the Coleman-De Luccia (CDL)
 instantons \cite{CL}.
 If this universe continues inflating
inside the bubble, then we obtain an open inflationary universe. Then by
a certain fine-tuning of parameters one can get any value of $\Omega_0$ in
the range $0< \Omega_0 < 1$ \cite{BGT,YSTapj}.

Even though the basic idea of this scenario was pretty simple, it was
very difficult to find a realistic open inflation model. The general
scenario proposed in \cite{BGT,YSTapj}  was based on investigation of
chaotic inflation and tunneling in the theories of a single scalar field
$\phi$. However, no models where this scenario could be successfully
realized have been proposed so far.
As it was shown in \cite{Open},
in the simplest models with polynomial potentials of the type of
${m^2\over 2} \phi^2-{\delta\over 3} \phi^3 + { \lambda\over 4} \phi^4$
the tunneling  occurs not  by bubble
formation, but by jumping onto the top of the potential barrier
described by the Hawking-Moss instanton \cite{HM}.   This process leads
to  formation of inhomogeneous domains of a
new phase, and the whole scenario fails. The main reason for this
failure is rather generic \cite{Toy}. Typically, CDL instantons exist
only  if $|\partial^2V| > H^2$
 during the
tunneling (here and in the rest of the paper $\partial^2V$ stays for
$\partial^2V/\partial\phi^2$). Meanwhile, inflation, which, according to
\cite{BGT,YSTapj},
begins immediately after the tunneling, typically requires
$|\partial^2V| \ll H^2$. These two conditions are almost incompatible.

This problem can be avoided in  models of two
scalar fields \cite{Open}.  However, in this paper we will concentrate
on the   one-field open inflation. We will remember  why it was so
difficult to realize this scenario. Then we will describe two models
where this can be accomplished; one of these models was proposed
recently in \cite{Toy}.  The main purpose of this paper is to
investigate   the CMB anisotropy in these models. As we will see, CMB
anisotropy in these models has some distinguishing features, which may
serve as a signature for the one-filed open inflation models.

\section{Toy models of one-field open inflation}\label{problems}

To explain the main features of the one-field open inflation models, let
us consider an effective potential $V(\phi)$ with a local minimum
at $\phi_0$, and a global minimum at $\phi=0$, where $V=0$.  In an
$O(4)$-invariant Euclidean spacetime with the metric
\begin{equation}\label{metric}
ds^2 =d\tau^2 +a^2(\tau)(d \chi_E^2+ \sin^2 \chi_E \,d \Omega_2^2)\ ,
\end{equation}
the scalar field $\phi$ and the three-sphere radius $a$ obey the
equations of motion
\begin{equation}\label{eq1}
\ddot\phi+3{\dot a\over a}\dot\phi
 =\partial V\ ,  ~~~
  \ddot a= -{8\pi \over 3} a (\dot\phi^2 +V) \ ,
\end{equation}
where dots denote derivatives with respect to $\tau$.
Here and in what follows we will use the units where
$M_p = G^{-1/2} =1$.

An instanton which describes the creation of an open universe
was first found by Coleman and De Luccia   \cite{CL}.  It is
given by a slightly distorted de~Sitter four-sphere of radius
$H^{-1}(\phi_0)$, with $a \approx H^{-1} \sin H \tau$. The field $\phi$
lies on the `true vacuum' side of the maximum of $V$ in a region near
$\tau = 0$, and it is very close to the
false vacuum, $\phi_0$, in the opposite part of the four-sphere near
$\tau_i  \approx \pi/H$, The scale factor $a(\tau)$ vanishes at the
points $\tau=0$ and $\tau=\tau_{\rm i}$. In order to get a
singularity-free solution, one must have $\dot\phi = 0$ and
$\dot a=\pm 1$ at $\tau=0$ and $\tau=\tau_{\rm i}$. This configuration
interpolates between some initial point $\phi_i \approx   \phi_0$ and
the final point $\phi_f$.  After an analytic continuation to the
Lorentzian regime, it describes an expanding   bubble  which  contains
an   open universe \cite{CL}.

Solutions of this type can exist only if the bubble can fit into   de
Sitter sphere of radius  $H^{-1}(\phi_0)$. To understand whether this
can happen, remember that at  small $\tau$ one has $a \sim \tau$, and
Eq. (\ref{eq1}) coincides with equation describing creation of a bubble
in Minkowski space, with $\tau$ being replaced by the bubble radius $r$:
\ $\ddot\phi+ {3 \over r}\dot\phi=\partial V$ \cite{Coleman}.
 Here the radius of the bubble can run from $0$ to $\infty$. Typically
 the bubbles have size greater than the Compton wavelength of the scalar
 field, $r \gtrsim m^{-1} \sim (\partial^2V)^{-1/2}$ \cite{nucl}.

In de Sitter space $\tau$ cannot be greater than ${\pi\over H}$, and in
fact the main part of the evolution of the field $\phi$ must end at
$\tau \sim {\pi \over 2 H}$. Indeed, once the scale factor reaches its
maximum at $\tau \sim {\pi \over 2 H}$, the coefficient
${\dot a\over a}$ in Eq. (\ref{eq1})  becomes negative, which
corresponds to anti-friction. Therefore if the field $\phi$  still
changes rapidly at $\tau > {\pi \over 2 H}$, it  experiences ever
growing acceleration near $\tau_{\rm f}$, and typically the solution
becomes singular \cite{HT}.
Thus the Coleman-De Luccia (CDL) instantons exist only if
${\pi \over 2 H} > (\partial^2V)^{-1/2}$, i.e. if $\partial^2V> H^2$.
This condition must be satisfied at small $\tau$, which corresponds to
the endpoint of the tunneling,  where inflation should begin in
accordance with the scenario of Ref. \cite{BGT,YSTapj}.
 But this condition is  opposite to the
standard inflationary condition $ \partial^2V\ll H^2$.

This means that immediately after the tunneling the field begins rolling
much faster than it was anticipated in \cite{BGT,YSTapj}. As a result,
in many models, such as the models with the effective potential
$V(\phi)={m^2\over2}\phi^2-{\delta\over3}\phi^3+{\lambda\over4}\phi^4$,
the open inflation scenario simply does not work \cite{Open,Toy}. This
problem is very general, and for a long time  we did not have any model
where this scenario could be realized.   We will describe two of these
models here, one of which was proposed recently in \cite{Toy}. We do not
know as yet whether it is possible to derive these models from some
realistic theory of elementary particles, so for the moment we consider
them simply as toy models of open inflation. Still we believe that these
models deserve investigation because they share the generic property of
all models of this class: As we expected. immediately after the
tunneling one has $\partial^2V> H^2$. As we will see, this condition
suppresses scalar perturbations of metric produced soon after the
tunneling. The supercurvature perturbations are also suppressed, whereas
the tensor perturbations in these models may be quite strong. These
features may help us to distinguish one-field models of open inflation
based on the Coleman-De Luccia tunneling from other models of open
inflation.

The first model which we are going to consider has the effective
potential of the following type:
\begin{equation}\label{toy}
V(\phi) = {m^2\phi^2\over 2}
 \left (1+{\alpha^2\over  \beta^2+(\phi -v)^2}\right).
\end{equation}
Here $\alpha$ $\beta$ and $v$ are some constants; we will assume that
$\beta \ll v$. The first term in this equation is the potential of the
simplest chaotic inflation model ${m^2\phi^2\over 2}$.  The second term
represents a peak of width $\beta$ with a maximum near $\phi = v$. The
relative hight of  this peak with respect to the potential
${m^2\phi^2\over 2}$ is determined by the ratio
${\alpha^2\over\beta^2}$.

\begin{figure}[Fig1]
 \hskip 1.5cm
\leavevmode\epsfysize=5cm \epsfbox{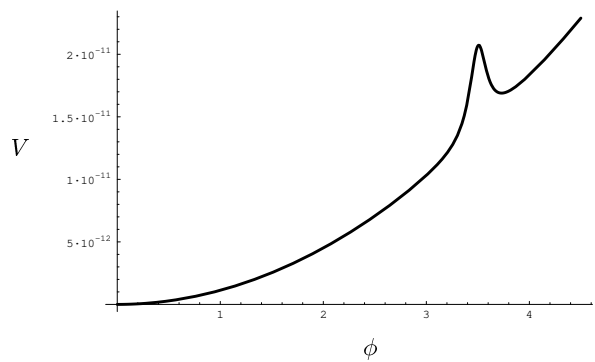}

\

\caption[Fig1]{\label{Pot} Effective potential our first  model, see
  Eq. (\ref{toy}). All values are given in units where $M_p = 1$.}

\end{figure}

As an example, we will consider the theory with $m = 1.5 \times
10^{-6}$, which is necessary to have a proper amplitude of density
perturbations during inflation in our model. We will take $v = 3.5$,
which, as we will see,  will provide about 65 e-folds of inflation after
the tunneling. By changing this parameter by few percent one can get any
value of $\Omega_0$ from $0$ to $1$.  For definiteness, in this section
we will take $ \beta^2 = 2 \alpha^2$, $\beta = 0.1$. This is certainly
not a unique choice; other values of these parameters to be considered
in the next section can also lead to a successful open inflation
scenario. The shape of the effective potential in this model is shown in
Fig. \ref{Pot}.

\begin{figure}[Fig1]
 \hskip 1.5cm
\leavevmode\epsfysize=10cm \epsfbox{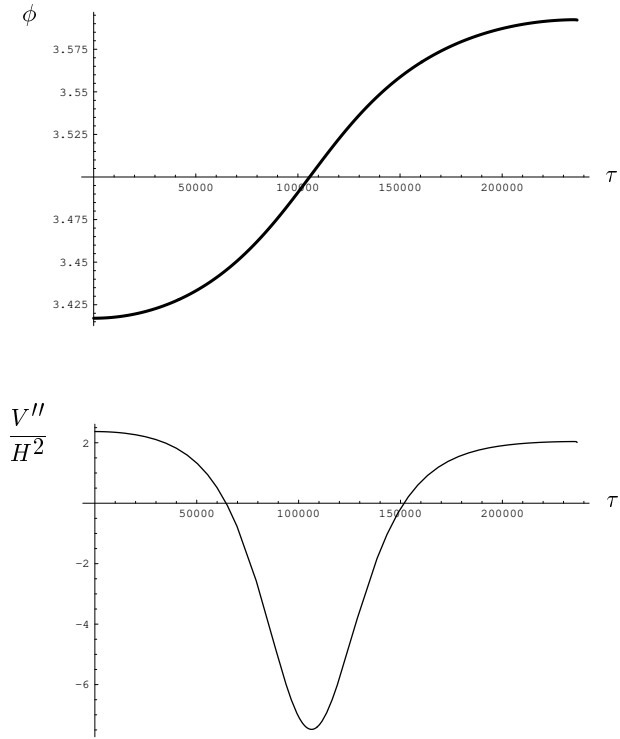}

\

\caption[Fig1]{\label{CDL} Coleman-De Luccia instanton in the first
  model. The upper panel shows the function $\phi(\tau)$, the lower
  panel demonstrates that most of the time during the tunneling one
  has $|\partial^2V| \gg H^2$.}

\end{figure}

As we see, this potential  coincides with ${m^2\phi^2\over 2}$
everywhere except a small vicinity of the point $\phi = 3.5$, but one
cannot roll from $\phi > 3.5$ to $\phi < 3.5$ without tunneling through
a sharp barrier. We have solved Eq. (\ref{eq1})  for this model
numerically and found that the Coleman-De Luccia instanton in this
model does exist. It is shown in Fig. \ref{CDL}.

The upper panel of Fig. \ref{CDL} shows the CDL instanton $\phi(\tau)$.
Tunneling occurs from $\phi_{i} \approx 3.6$ to
$\phi_{f} \approx 3.4$. The energy density decreases in this process,
$V(\phi_f) < V(\phi_i)$. The lower panel of Fig. \ref{CDL} shows the
ratio $\partial^2V/H^2$. Almost everywhere along the instanton
trajectory $\phi(\tau)$ one has $|\partial^2V| > H^2$. That is exactly
what we have expected on basis of our general arguments concerning CDL
instantons.

 An interesting feature of the CDL instantons is that the evolution of
 the field $\phi$ does not begin  exactly  at the local minimum of the
 effective potential.    This is similar to what happens in  the
 Hawking-Moss  case  \cite{HM}, where tunneling begins and ends not at
 the local minimum but at the top of the effective potential; see
 \cite{ALOpen} for a recent discussion of this issue. This
 unconventional feature of the CDL instantons was not emphasized in
 \cite{CL} because the authors concentrated on the thin wall
 approximation where this effect disappears. For a proper interpretation
 of these instantons, just as in the Hawking-Moss case,  one may either
 glue to the point $\tau_f$ a de Sitter hemisphere corresponding to the
 local minimum of the effective potential \cite{ALOpen}, or use a
 construction proposed in \cite{BC}.  It would be very desirable to
 verify the Coleman-De Luccia approach by a complete Hamiltonian
 analysis of the tunneling in inflationary universe.

\begin{figure}[Fig1]
 \hskip 1.5cm
\leavevmode\epsfysize=5cm \epsfbox{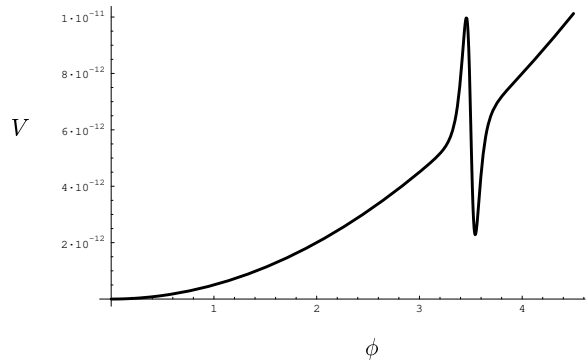}

\

\caption[Fig1]{\label{Pot2} Effective potential in our second model,
  Eq. (\ref{toy2}). All values are given in units where $M_p = 1$.}

\end{figure}

The second model   has the effective potential of the following type:
\begin{equation}\label{toy2}
V(\phi) =    {m^2\over 2} \left (\phi^2+ B^2{ \sinh A(\phi-v) \over
    \cosh^2 A(\phi-v)  }\right)
\end{equation}
Here $A$, $B$ and $v$ are some constants. As an example, we will consider
the theory with $m = 1.0 \times 10^{-6}$,  $v = 3.5$,  $A =20$, and
$B=4$. The shape of the effective potential in this model is shown in
Fig. \ref{Pot2}.

\begin{figure}[Fig1]
 \hskip 1.5cm
\leavevmode\epsfysize=10cm \epsfbox{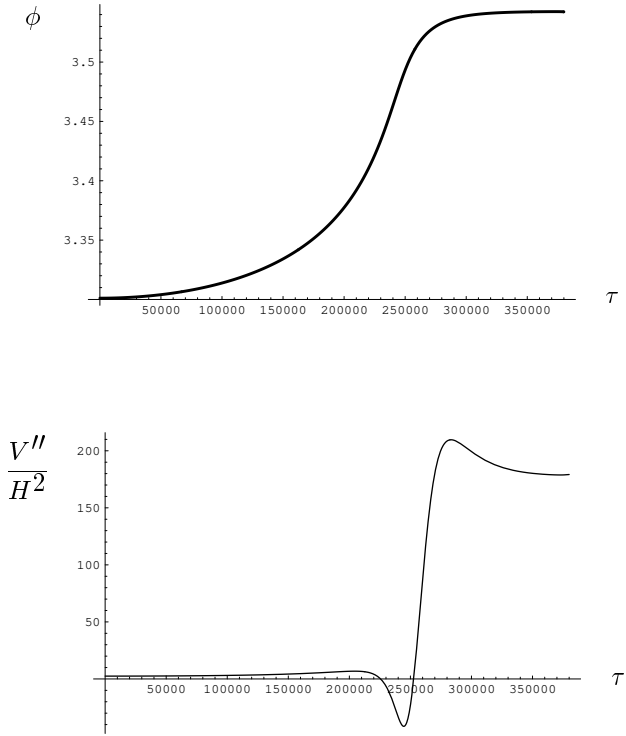}

\

\caption[Fig1]{\label{CDL2} Coleman-De Luccia instanton in our second
  model. The upper panel shows the function $\phi(\tau)$, the second one
  shows the ratio $|\partial^2V|/H^2$, which remains very large during
  the tunneling.}

\end{figure}

The Coleman-De Luccia instanton in this model   is shown in
Fig. \ref{CDL2}. The upper panel of Fig. \ref{CDL2} shows the  instanton
$\phi(\tau)$.  Tunneling occurs from $\phi_{i} \approx 3.54$, which
almost exactly coincides with the position of the local minimum of
$V(\phi)$,  to $\phi_{f} \approx 3.30$. The energy density increases in
this process, $V(\phi_f) > V(\phi_i)$. This may seem unphysical, but in
fact such jumps are possible because of the gravitational effects. A
similar effect occurs during the  Hawking-Moss tunneling to the local
maximum of the effective potential \cite{HM}. The lower panel of
Fig. \ref{CDL2} shows that almost everywhere along the trajectory
$\phi(\tau)$ one has $|\partial^2V| \gg H^2$.   After the tunneling  the
scalar field slowly rolls down and then   oscillates near the minimum of
the effective potential at $\phi = 0$. During the stage of the slow
rolling, the scale factor in the models which we investigated expands
approximately $e^{65}$ times.

\section{CMB anisotropy in the open inflation models}
Just as we expected, in both models the tunneling brings the field to
the region where  $|\partial^2V| > H^2$. Therefore the usual scalar
perturbations of density are not produced in these models immediately
after the open universe formation. As we will see now, this leads to a
suppression of the contribution of these perturbations to the CMB
anisotropy at $\ell \lesssim 10$.

In addition to these perturbations, we could encounter supercurvature
perturbations which are produced in the false vacuum outside the bubble
and may later penetrate into its interior during the bubble
expansion. However, we did not find any supercurvature perturbations in
these models.

The reason why there are no supercurvature perturbations in the second
model is pretty simple: The curvature of the effective potential in the
false vacuum is much greater than $H^2$, so these perturbations are not
produced outside the bubble.

For the first model the reason for the absence of the supercurvature
modes is less obvious because in the false vacuum there one has
$\partial^2V \ll H^2$. However, all information about the interior of
the bubble can be obtained by the analytical continuation of the CDL
instanton, which begins away from the false vacuum, in a state with
$\partial^2V > H^2$. The fact that there is no region where
$\partial^2V\ll H^2$ in the CDL instanton implies that the initial
distance from the center of the bubble to the place where $\partial^2V$
becomes smaller than $H^2$ (in the false vacuum outside of the CDL
instanton) is greater than $2H^{-1}$, i.e., it is greater than twice the
size of the event horizon in de Sitter space.
As a result, the fluctuations produced in the false vacuum do not
penetrate into the bubble.

In addition to the scalar perturbations, there also exist tensor
perturbations. Unlike the standard inflation scenario, it is known that
the fluctuations of the bubble wall contribute to the low frequency
spectrum of tensor perturbations and the contribution can dominate over
the scalar spectrum \cite{STYtensor,Bell}. In fact, we shall see that they
can be quite significant and dominate the CMB anisotropy spectrum for
small $l$.

Below we present the scalar and tensor spectra for three models:
Two of them are those discussed in the previous section. The third model
is the one with the same potential form as the first model but with
a different value of $\beta$; $\beta^2=\alpha^2/2=0.0025$.
To compute the spectra, we adopt a gauge-invariant method developed by
Garriga, Montes, Sasaki and Tanaka \cite{GMST1,GMST2}.  Then we show the
resulting CMB anisotropy spectra on large angular scales.

\subsection{Scalar and tensor perturbation spectra}
Let us first summarize the procedure to obtain the scalar and tensor
spectra. The metric describing the Lorentzian bubble configuration is
given by the analytic continuation of (\ref{metric})
with $\chi_E=-i\chi_C+\pi/2$:
\begin{eqnarray}
  \label{Cmetric}
  ds^2 =d\tau^2 +a^2(\tau)(-d\chi_C^2+ \cosh^2 \chi_C \,d \Omega_2^2) \,.
\end{eqnarray}
The scalar field configuration is still given by $\phi=\phi(\tau)$.
In the one-field models of one-bubble open inflation, the scalar
perturbation is conveniently described by a variable $\bbox{q}$, which
is essentially equivalent to the gravitational potential perturbation
$\Psi_N$ in the Newton gauge,
\begin{eqnarray}
  \label{qdef}
  \bbox{q} ={\Psi_N\over 4\pi G\dot\phi}\,.
\end{eqnarray}
Here and below we recover $G$ in equations.
The (even parity) tensor perturbation is described by a variable  
$\bbox{w}$,
whose relation to the transverse-traceless metric perturbation in the
open universe will be given later. There are also odd parity modes for
the tensor perturbation. But since the odd parity modes do not
contribute to the CMB anisotropy, we shall not discuss them.
Here we just mention that the form of the
Lagrangians for both $\bbox{q}$ and $\bbox{w}$ is that for a scalar
field with $\tau$-dependent mass \cite{GMST1,GMST2}.

We quantize the variables $\bbox{q}$ and $\bbox{w}$ on the
$\chi_C={\rm const.}$ hypersurface which is a Cauchy surface and which
contains all the information of the bubble configuration.
We expand them in terms of the spherical harmonics
$Y_{\ell m}$ and spatial eigenfunctions $\bbox{q}^p$ and
$\bbox{w}^p$ with eigenvalue $p^2$:
\begin{eqnarray}
  \bbox{q}&=&\sum \hat a_{p\ell m}f^{p\ell}(\chi_C)
    \bbox{q}^p(\tau)Y_{\ell m}(\Omega_2) +{\rm h.c.}\,,
\label{qexpand}\\
  \bbox{w}&=&\sum \hat b_{p\ell m}f^{p\ell}(\chi_C)
    \bbox{w}^p(\tau)Y_{\ell m}(\Omega_2) +{\rm h.c.}\,,
\label{wexpand}
\end{eqnarray}
where $\hat a_{p\ell m}$ and $\hat b_{p\ell m}$ are the annihilation
operators.
The spatial eigenfunctions $\bbox{q}^p$ and $\bbox{w}^p$ satisfy,
respectively,
\begin{eqnarray}
\left[-{d^2\over d\eta_C^2}+U_{S}(\eta_C)\right]{\bbox{q}}^p
    &&=p^2{\bbox{q}}^p\,;
\nonumber\\
 U_{S}=&&{4\pi}G{\phi'}^2 +\phi'\left({1\over \phi'}\right)''-4\,,
\label{qeq}\\
\left[-{d^2\over d\eta_C^2}+U_{T}(\eta_C)\right]{\bbox{w}}^p
    &&=p^2{\bbox{w}}^p\,;
\nonumber\\
 U_{T}=&&{4\pi}G{\phi'}^2 \,,
\label{weq}
\end{eqnarray}
where $d\eta_C=d\tau/a(\tau)$ and primes denote derivatives with respect
to $\eta_C$. The potentials $U_S$ and $U_T$ both vanish for
$\eta_C\to\pm\infty$, but $U_S$ is not necessarily positive definite. It
then follows that if there exists a bound state for this eigenvalue
equation, it exists discretely at some $p^2<0$ and corresponds to a
supercurvature mode of the scalar spectrum. On the other hand, $U_T$ is
manifestly positive definite and there is no supercurvature mode in the
tensor spectrum. For both scalar and tensor perturbations, the spectrum
is continuous for $p^2>0$. As noted before, we found no
supercurvature mode in all of the three models.

The equation for $f^{p\ell}$ turns out to be
model-independent and is given by
\begin{eqnarray}
\label{feq}
&& \left[-{1\over\cosh^2\chi_C}{\partial\over \partial\chi_C}
    \cosh^2\chi_C{\partial\over \partial\chi_C}
    -{\ell(\ell+1)\over \cosh^2\chi_C}\right] f^{p\ell}
\nonumber\\
&&\qquad  =(p^2+1) f^{p\ell}.
\end{eqnarray}
In accordance with the Euclidean approach to the tunneling, we take the
quantum states of $\bbox{q}$ and $\bbox{w}$ to be the Euclidean  
vacua. This
implies that the positive frequency function $f^{p\ell}$ is regular at
$\chi_E=\pi/2$ ($\chi_C=0$). Apart from the normalization, the solution
is
\begin{equation}
 f^{p\ell}(\chi_C)\propto{1\over\sqrt{\cosh \chi_C}}
   P^{-\ell-1/2}_{ip-1/2} (i\sinh \chi_C),
\end{equation}
where $P^{\mu}_{\nu}$ is the associated Legendre function of the first
kind. The normalizations of the mode functions $f^{p\ell}\bbox{q}^p$
and $f^{p\ell}\bbox{w}^p$ are determined by the standard Klein-Gordon
normalization of a scalar field.

We then analytically continue $\bbox{q}$ and $\bbox{w}$ to the region
just inside the lightcone emanating from the center of the bubble,
i.e., to the region of the open universe, by $\chi=\chi_C+i\pi/2$ and
$t=-i\tau$ (or $\eta=\eta_C-i\pi/2$). The metric there is given by
\begin{eqnarray}
  \label{Ometric}
  ds^2 &=&-dt^2 +a^2(t)(d\chi^2+ \sinh^2 \chi \,d \Omega_2^2)
\nonumber\\
 &=&a^2(\eta)(-d\eta^2+d\chi^2+\sinh^2\chi\,d\Omega_2^2) \,.
\end{eqnarray}
Then the function $f^{p\ell}$ just becomes the radial function of a
spatial harmonic function on a unit spatial 3-hyperboloid,
\begin{equation}
\left(\mathop{\Delta}^{(3)}+p^2+1\right)Y^{p\ell m} =0\,;\quad
Y^{p\ell m}=f^{p\ell}(\chi)Y_{\ell m}(\Omega_2).
\label{laplace}
\end{equation}
On the other hand, the spatial eigenfunctions $\bbox{q}^p$ and
$\bbox{w}^p$ become the temporal mode functions for the scalar and
tensor perturbations, respectively, in the open universe. Note that
$p\sim1$ corresponds to the comoving spatial curvature scale.
The evolution equations for $\bbox{q}^p$ and $\bbox{w}^p$ take the  
same forms
as Eqs.~(\ref{qeq}) and (\ref{weq}), respectively, with the replacement
$\eta_C\to\eta$.
We solve Eqs.~(\ref{qeq}) and (\ref{weq}) until the scale of the
perturbation is well outside the Hubble horizon scale, i.e., until
\begin{eqnarray}
  \label{superhorizon}
a^2H^2\gg p^2+1\,.
\end{eqnarray}
Here and in what follows, $H$ is not the inverse of the de Sitter radius
but $H=\dot a/a$.

The important quantity that determines the primordial density
perturbation spectrum as well as the large angle scalar CMB anisotropies
is the curvature perturbation on the comoving hypersurface,
${\cal R}_c$.
The comoving hypersurface is the one on which the scalar field
fluctuation $\delta\phi$ vanishes. It is related to $\bbox{q}$ as
\begin{equation}
  \label{calRc}
  {\cal R}_c^p=-{4\pi G \dot\phi\,\bbox{q}^p}
  +{H\over a\,\dot\phi^2}
    {d\over dt}\left(a\,\dot\phi\,\bbox{q}^p\right).
\end{equation}
Just as in the case of the flat universe inflation, ${\cal R}_c$ remains
constant in time until the perturbation scale re-enters the Hubble
horizon \cite{YST}.

On the other hand, the even parity tensor perturbation in the
open universe is described as
\begin{eqnarray}
  \label{eptensor}
  \delta g_{ij}=a^2 t_{ij}\,;\quad
 t_{ij}=\sum \hat b_{p\ell m}U_p(\eta)Y^{(+)p\ell m}_{ij}+\hbox{h.c.}\,,
\end{eqnarray}
where $Y^{(+)p\ell m}_{ij}$ are the even parity tensor harmonics on the
unit 3-hyperboloid \cite{Tomita}. After an appropriate choice of the
normalization factor, $U_p$ is given in terms of
$\bbox{w}^p$ as \cite{GMST2}
\begin{eqnarray}
  \label{Up}
  U_p=-{8\pi G\over a\,(p^2+1)}{d\over dt}\,(a\,\bbox{w}^p).
\end{eqnarray}
Similar to the case of the scalar perturbation, $U_p$ is known to remain
constant in time on superhorizon scales.

\begin{figure}[Fig]
\leavevmode\epsfysize=8cm \epsfbox{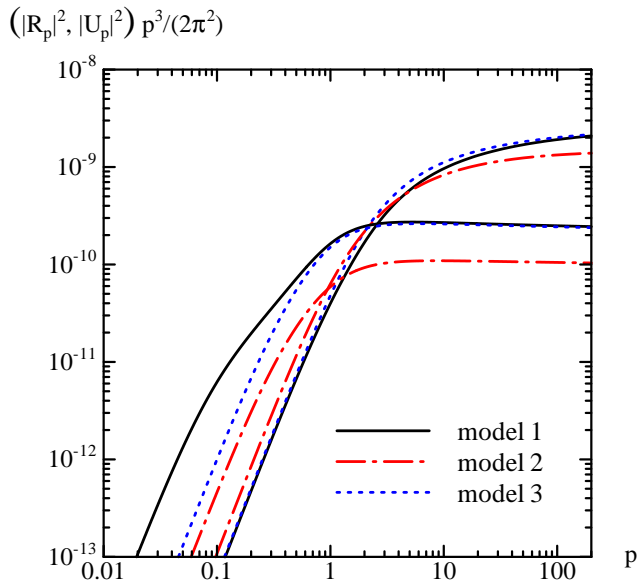}

\

\caption[Fig]{\label{Nspec} The spectra of scalar and tensor
  perturbations (per logarithmic interval of $p$) for Models 1,
  2 and 3. The three curves that gradually increase as $p$ are the
  scalar spectra, and the other three that level off at large $p$ are
  the tensor spectra. The spectra for Models 1, 2 and 3 are shown by
  the solid, dot-dashed and dotted curves, respectively.}

\end{figure}

In Fig.~\ref{Nspec}, the scalar and tensor perturbation spectra for the
first, second and third models (which we call Models 1, 2 and 3,
respectively) are shown. Let us recall their model parameters:
\begin{eqnarray*}
 \hbox{Model 1:}~~&&\hbox{Eq.~(\ref{toy}) with}
     ~\alpha^2=0.005, ~\beta^2=2\alpha^2,
\\
&&\quad v=3.5, ~ m=1.5\times10^{-6}.
\\
 \hbox{Model 2:}~~&&\hbox{Eq.~(\ref{toy2}) with}
     ~A=20, ~B=4,
\\
&&\quad v=3.5, ~ m=1.0\times10^{-6}.
\\
 \hbox{Model 3:}~~&&\hbox{Eq.~(\ref{toy}) with}
     ~\alpha^2=0.005, ~\beta^2=\alpha^2/2,
\\
&&\quad v=3.5, ~ m=1.5\times10^{-6}.
\end{eqnarray*}

First let us consider the scalar spectra. As mentioned previously,
there are no supercurvature modes in the present models. So the scalar
perturbations are completely described by the continuous spectra shown
in Fig~\ref{Nspec}.
As seen from the figure, the scalar spectra for the three models are all
alike: On the low frequency end, they decrease sharply as $p$ decreases,
while they gradually increase for $p\agt10$. As discussed in the
previous section, one can interpret this feature as due to the common
evolutionary behavior of any successful one-field model with the CDL
tunneling. The scalar field evolves rapidly for the first few expansion
times when $\partial^2V>H^2$ and eventually decelerates as the slope of
the effective potential becomes flatter. For $p\gg1$, the spectrum
approaches the one given by the standard formula for the flat universe
inflation models.
As shown in \cite{Toy}, the gradual increase gives rise to a peak in the
spectrum at $p\sim10^4$, which may have significant implications to the
structure formation in the universe.

To understand the shape of the scalar spectrum more quantitatively,
it is useful to compare the computed spectrum with the following
analytic formula \cite{YST,GMST2},
\begin{eqnarray}
  \label{analyticRc}
  |{\cal R}_c^p|^2{p^3\over2\pi^2}=
\left({H^2\over2\pi\dot\phi}\right)^2_{t=t_p}
{\cosh\pi p+\cos\delta_p\over\sinh\pi p}{p^2\over1+p^2}\,,
\end{eqnarray}
where $t_p$ is an epoch slightly after the perturbation scale goes out
of the Hubble horizon. This formula assumes $\partial^2V\ll H^2$ and the
slow time variation of $\partial^2V$. The angle $\delta_p$  
describes the effect of the bubble wall, which
is known to behave as $\delta_p-\pi\propto p$ for $p\to 0$. The low
frequency part of the spectrum is most suppressed when $\delta_p=\pi$.
This case corresponds to the case when $\partial^2V\gg H^2$ on
the false vacuum side of the instanton \cite{YST}.

In our case the condition $\partial^2V\ll H^2$ is violated at the  
first stages
after the bubble formation. Therefore Eq.~(\ref{analyticRc})
should be somewhat modified for small $p$. Indeed, fluctuations with  
small $p$ are produced soon after the tunneling. But immediately after
the tunneling one has $\partial^2V > H^2$ in all models where the
Coleman-De Luccia instantons exist.  Therefore the perturbations with
the wavelength greater than $H^{-1}$ will not become ``frozen''
immediately after the tunneling. They
will freeze somewhat later, when the field $\phi$ will roll to the  
area with
$\partial^2V\ll H^2$. But at that time their wavelength increases  
and their
amplitude becomes smaller. As a result, Eq.~(\ref{analyticRc})
provides a
good description of the spectrum at large $p$, but at small $p$ the  
amplitude of perturbations will be somewhat smaller than that given by
Eq.~(\ref{analyticRc}). This expectation is confirmed by the results of
our numerical investigation.

\begin{figure}[Fig]
\hskip 1.5cm
\leavevmode\epsfysize=8cm \epsfbox{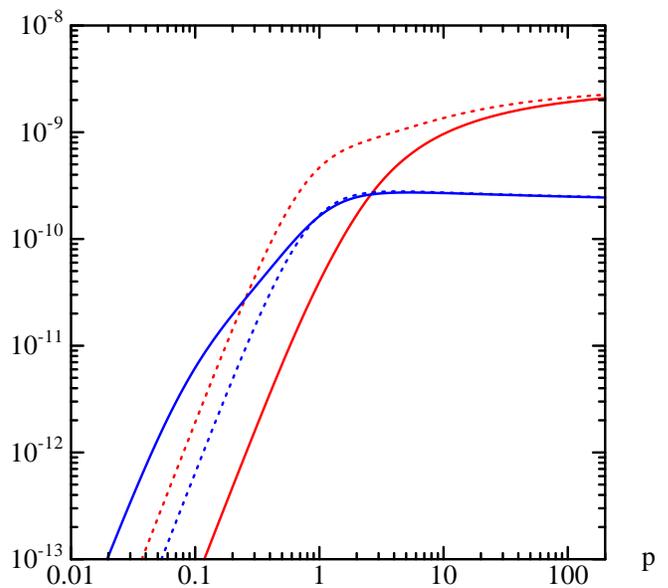}

\

\caption[Fig]{\label{Aspec} Comparison of the scalar and tensor spectra
  of Model 1 with the ones given by analytic formulas
  (\ref{analyticRc}) and (\ref{analyticUp}), respectively.
  The upper and lower dotted lines show the scalar and tensor formulas,
  respectively. The solid lines that coincide with the upper and lower
  dotted lines for $p\gg1$ are, respectively, the computed scalar and
  tensor spectra.}

\end{figure}

In Fig.~\ref{Aspec}, this comparison is made for Model 1. In the figure,
the upper dotted line shows the formula (\ref{analyticRc}) and the solid
line that approaches it for $p\gg1$ is the computed one. We choose $t_p$
to be the time when $a^2(2H^2-\partial^2V)=2(1+p^2)$ and
$\delta_p=\pi$. As one can see, the computed spectrum at small $p$ is
significantly more suppressed than the most suppressed case of the
analytic formula. As we shall see below, this large suppression relative
to the analytic formula causes a large suppression of the CMB anisotropy
at small $\ell$.
The suppression of scalar perturbations with small $p$ and $\ell$  and the
absence of supercurvature perturbations seem to be a generic property of
the models of one-field open inflation based on the CDL tunneling.
On the other hand, the spectrum become almost indistinguishable from
the one given by Eq.~(\ref{analyticRc}) for $p\gg1$. Thus the tilt of
the spectrum (with a positive power-law index) is due to the slowing
down of the evolution of $\phi$.

Now let us consider the tensor spectra.
The spectra for Models 1 and 3 are indistinguishable at $p\agt1$,
while the spectrum for Model 2 is about a factor of 2 smaller.
This difference is due to the difference in the choice of the
mass parameter: The mass square for Models 1 and 3 is $1.5^2=2.25$
greater than that for Model 2. This results in the difference in
$H^2$. In fact, if we multiply the spectrum of Model 2 by 2.25, it
becomes almost indistinguishable from the spectrum of Model 3 for the
whole range of $p$.
Turning to the low frequency behavior, the spectrum of Model 1 at
$p\alt1$ differs considerably from that of Model 3: The former is
larger by an order of magnitude relative to the latter at small
$p$. This enhancement is due to the wall fluctuation modes. Recall that
the parameter $\beta$ for Model 1 is larger than that for Model 3. Since
a larger $\beta$ means a lower potential barrier, the wall tension is
smaller for Model 1 than for Model 3. This makes the wall of Model 1
easier to vibrate.

A non-dimensional quantity that represents the strength of the wall
tension is given by the following integral over the instanton background
\cite{STYtensor}:
\begin{eqnarray}
  \label{Deltas}
  \Delta s =4\pi G\int\phi'{}^2d\eta_C\,.
\end{eqnarray}
For Models 1, 2 and 3, the values of $\Delta s$ are found as
\begin{eqnarray}
  \label{Dsvalue}
  \hbox{Model 1:}~~&& \Delta s=0.1681\,, \nonumber\\
  \hbox{Model 2:}~~&& \Delta s=0.6614\,, \nonumber\\
  \hbox{Model 3:}~~&& \Delta s=0.6640\,.
\end{eqnarray}
In the thin-wall limit, $\Delta s=4\pi GR_WS_1$, where
$R_W$ is the wall radius and $S_1$ is the surface tension. Further,
in this limit, $\Delta s$ is always smaller than unity and the low
frequency spectrum is enhanced by a factor $\sim1/\Delta s^{2}$ for the
width $\Delta p\sim\Delta s$ \cite{STYtensor}. In the present case, as
we have seen in section~\ref{problems}, the bubble walls are not at all
thin. Nevertheless, this qualitative feature expected from the thin-wall
limit is in good agreement with the computed tensor spectra.

To see the effect of wall fluctuations more clearly, in
Fig.~\ref{Aspec}, the tensor spectrum for Model 1 is compared with
that given by the following approximate analytic formula derived in
\cite{TStensor,Bucher,GMST1,GMST2}:
\begin{eqnarray}
  \label{analyticUp}
  |U_p|^2{p^3\over2\pi^2}=32\pi G
\left({H\over2\pi}\right)^2_{t=t_p}
{\cosh\pi p-1\over\sinh\pi p}{p^2\over1+p^2}\,,
\end{eqnarray}
where we took the large tension limit, which makes
the wall fluctuations least effective. As seen from Fig.~\ref{Aspec},
the analytic formula agrees very well with the computed spectrum for
$p\agt1$. Hence the difference at $p\alt1$ is totally due to the wall
fluctuation modes.
If one compares the analytic tensor spectrum in Fig.~\ref{Aspec} with
the tensor spectrum of Model 3 in Fig.~\ref{Nspec}, one sees they almost
coincide with each other. This is in accordance with the fact that
$\Delta s$ of Model 3 is large, as shown in Eq.~(\ref{Dsvalue}).
Thus the bubble wall fluctuations are highly suppressed in Model 3
(and in Model 2) due to the large wall tension.

\subsection{Large angle CMB spectra}

We now discuss the CMB anisotropies for Models 1, 2 and 3.
We focus on the CMB anisotropy spectrum for $\ell\le20$. Since the
contribution of scalar perturbations is dominated by the effect
of gravitational potential perturbations, we take account of only the
so-called Sach-Wolfe and integrated Sach-Wolfe effects. Although there
is a possibility that $\Omega_0$ is dominated by $\Omega_\Lambda$, here 
we assume the present universe is matter-dominated;
$\Omega_0=\Omega_{\rm matter}$.

Before going into discussion, we note one subtlety.
In the one-bubble open universe scenario, the duration of inflation
inside the bubble is directly related to the value of $\Omega_0$
today. In other words, once the model parameters are fixed, the
duration of inflation is fixed and consequently so the value of
$\Omega_0$. However, $\Omega_0$ depends rather sensitively on the values
of the model parameters. In particular, it takes a very small change in
$v$ to give a different $\Omega_0$. But such a change will not cause a
change in the shape of perturbation spectra. Furthermore, the efficiency
of reheating (or preheating) at the end of inflation will also affect
the value of $\Omega_0$. So, depending on a grand scenario one has in
mind, the resulting $\Omega_0$ will be different. Because of these
reasons, below we present the CMB anisotropies of Models 1, 2 and 3 for
several different values of $\Omega_0$ by artificially varying it.

\begin{figure}[Fig]
\hskip 1.5cm
\leavevmode\epsfysize=8cm \epsfbox{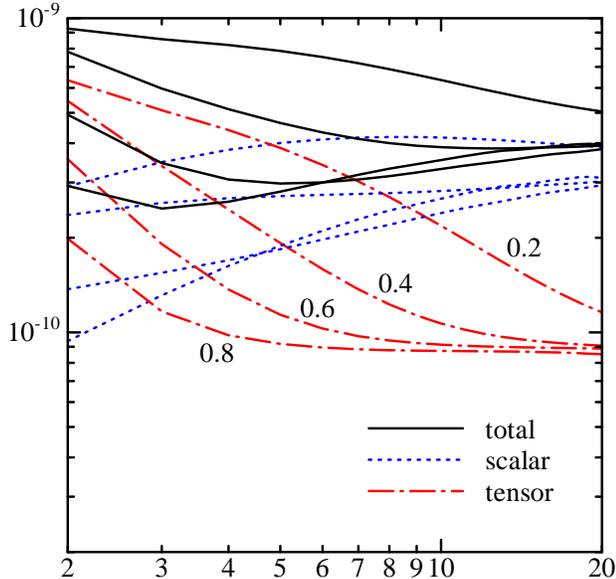}

\

\caption[Fig]{\label{CMB1} The CMB spectra of Model 1 for several values
  of $\Omega_0$. The solid lines show the total CMB spectra. The dotted
  lines show the scalar contributions and the dot-dashed lines the
  tensor contributions. The values of $\Omega_0$ are 0.2, 0.4, 0.6 and
  0.8 from top to bottom (at $\ell=3$) for each kind of curves.}

\end{figure}

\begin{figure}[Fig]
\hskip 1.5cm
\leavevmode\epsfysize=8cm \epsfbox{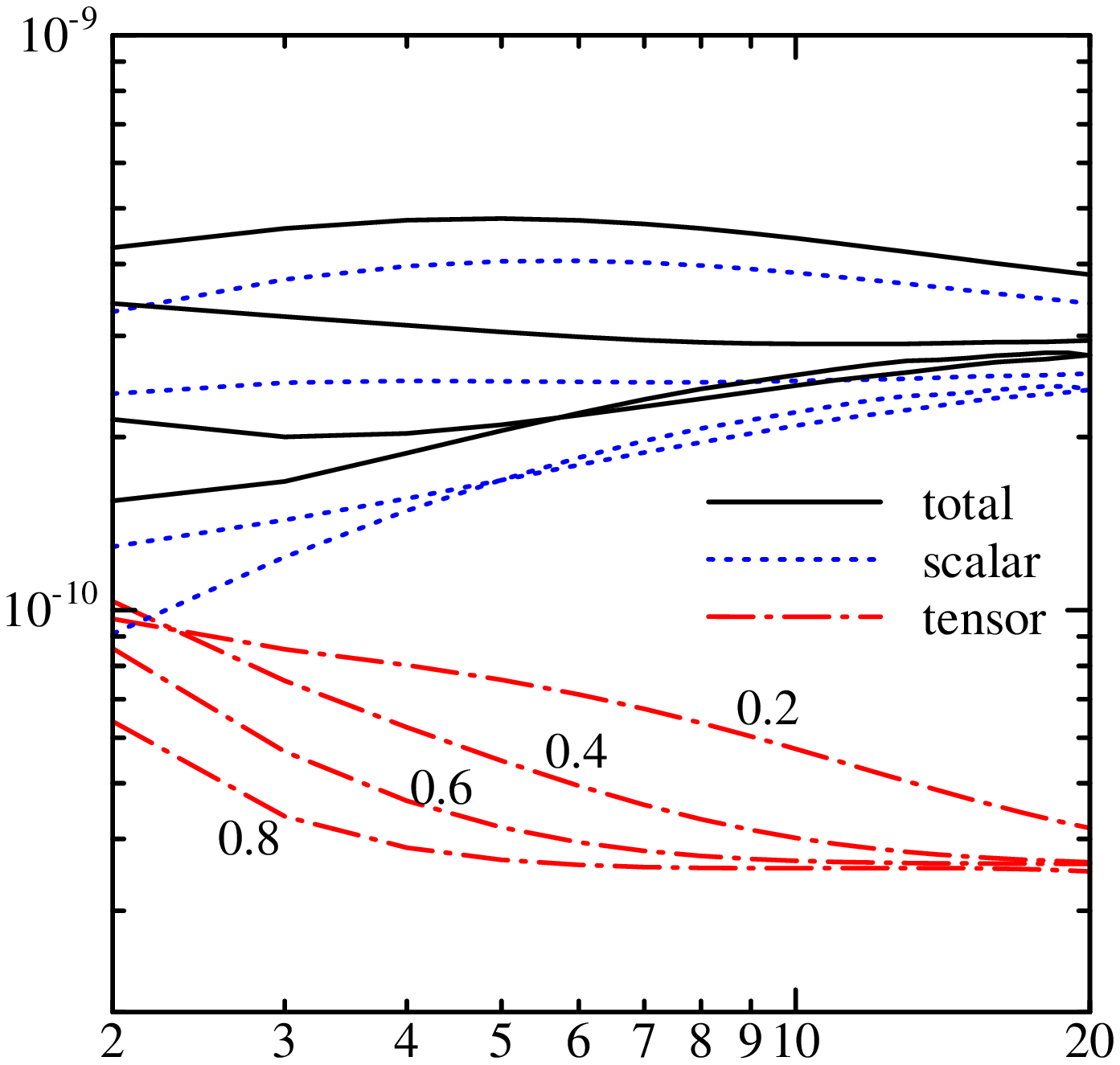}

\

\caption[Fig]{\label{CMB2} The same as Fig.~\ref{CMB1}, but for Model
  2.}

\end{figure}

\begin{figure}[Fig]
\hskip 1.5cm
\leavevmode\epsfysize=8cm \epsfbox{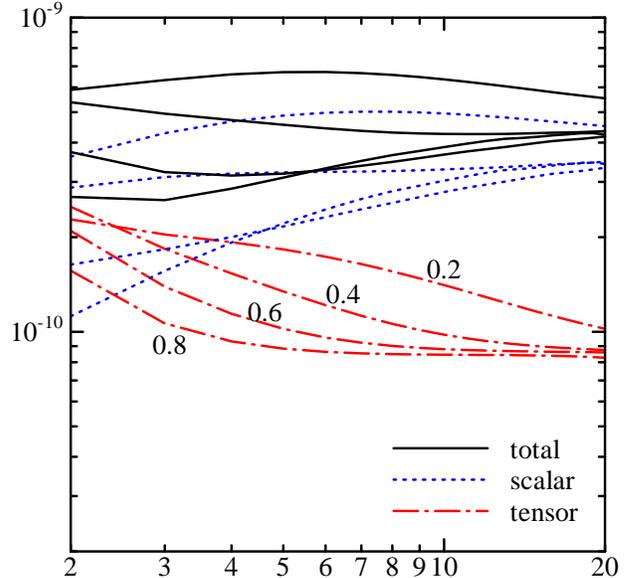}

\

\caption[Fig]{\label{CMB3} The same as Fig.~\ref{CMB1}, but for Model
  3.}

\end{figure}

The computed CMB spectra $\ell(\ell+1)C_\ell$ for Models 1, 2 and 3
are shown in Figs.~\ref{CMB1}, \ref{CMB2} and \ref{CMB3},
respectively. The amplitudes shown there are the absolute amplitudes of
the spectra for the given parameter values. It should be noted, however,
that the amplitude can be tuned to fit the observed value (at certain
$\ell$) by changing the value of $m$ if necessary. So, the important
point is the relative amplitudes of the scalar and tensor contributions
and their spectral shapes.

The scalar CMB anisotropies show similar spectral behavior for all the
models. Namely, their amplitudes are suppressed at small $\ell$.
This behavior is due to the large suppression of the scalar spectra at
$p\alt 10$ mentioned in the previous subsection. If one compares the
present results with the ones shown in Figs.~4, 5 and 6 of \cite{YSTapj},
one sees that the tendency is opposite: The scalar spectra obtained in
\cite{YSTapj} have a feature that they gradually decrease as $\ell$
increases. This is due to the integrated Sach-Wolfe effect and it is
usually what one expects for open universe models. On the contrary, in
the present case, because of the large suppression of the scalar
spectra at $p\alt10$, the corresponding CMB spectra increase for
increasing $\ell$ and level off around $\ell\sim10$.

As expected from the tensor perturbation spectra shown in
Fig.~\ref{Nspec}, the tensor CMB anisotropies at $\ell\alt 5-10$ are
large in Model 1 due to large wall fluctuations, while they are
small in Models 2 and 3.
For Model 1, this enhancement causes a rise in the total spectra for
$\ell\alt 5$, which does not seem to fit with the observed spectrum by
COBE-DMR \cite{COBE}. On the other hand, the tensor contribution to the
CMB anisotropies of Models 2 and 3 is small. As a result, the total
spectra of Models 2 and 3 turn out to be rather flat, which is
consistent with the COBE spectrum.

\section{Conclusions}

Despite a lot of progress in our understanding of various versions of open
inflation, until now we did not know how the spectrum of CMB may  
look in the
simplest one-field open inflation models. Previous calculations have  
been based
on the assumption that   the usual inflationary perturbations are produced
inside the bubble immediately after it is formed.

However, as we have argued (see also \cite{Toy}), bubbles appear only if
$\partial^2V > H^2$ at the moment of their formation in the  
one-field models.
This means that the usual inflationary perturbations are not  
produced at that
time.

In this paper we have studied the spectrum of CMB in several  
different models
of one-field open inflation. At $\ell\gg10$ the spectrum coincides  
with the
spectrum obtained in the earlier papers on open inflation, since the  
mechanism
of the bubble production is not very important for the behavior of the
perturbations   on  scale much smaller than the size of the bubble.  
The main
difference in the spectrum of CMB occurs at $\ell\lesssim O(10)$.

We have found that the spectrum of scalar CMB anisotropies has a  
minimum at
small $\ell$, and reaches a plateau at $\ell = O(10)$. The existence  
of this
minimum is a model-independent feature of the spectrum related to  
the fact that
$\partial^2V > H^2$ at the moment of the bubble formation in the one-field
models. In all models which we have studied  there are no supercurvature
perturbations. Tensor CMB anisotropies are peaked at $\ell=2$. Relative
magnitude of the scalar CMB spectra versus
tensor CMB spectra at small $\ell$ depends on the parameters of the
models, and in particular on the value of $\Omega_0$. 
In some of the models, tensor perturbations are too large,
which rules these models out. This effect is especially pronounced in the
models with $\Omega_0 \ll 1$. In some other models the tensor  
perturbations are
very small even for  $\Omega_0 \ll 1$, and the combined spectrum of  
perturbations has a minimum at small $\ell$. In future satellite
missions one could measure the tensor spectrum
via polarization. This  would make it possible to  identify the scalar and tensor contributions to the CMB anisotropy \cite{Bell} and to compare them with the predictions of the one-field models of open inflation.   We conclude that the the
spectrum of CMB in one-field models of open inflation has certain
features which will help us to
verify these models and to distinguish them from other versions of  
inflationary theory.

\subsection*{Acknowledgments}

It is a pleasure to thank J. Garc\'{\i}a--Bellido and R. Bousso for
useful and stimulating discussions. The work of A.L.  was supported in
part by NSF grant PHY-9870115, and the work of M.S. and T.T. was
supported in part by Monbusho Grant-in-Aid for Scientific Research
No.~09640355.


\begin{thebibliography}{999}
\bibitem{CL} S. Coleman and F. De Luccia,   {  Phys. Rev.} {\bf  
D21},  3305
(1980).

\bibitem{Gott}    J.R. Gott,
{ Nature} {\bf 295}, 304 (1982); J.R. Gott, and T.S. Statler, {  
Phys. Lett. }
{\bf 136B}, 157 (1984).

 \bibitem{BGT} M. Bucher, A.S. Goldhaber, and N. Turok,
 Phys. Rev. {\bf D52}, 3314  (1995).

 \bibitem{YSTapj} K. Yamamoto, M. Sasaki and T. Tanaka,
  Astrophys. J. {\bf 455}, 412 (1995).

 \bibitem{Open} A.D. Linde,     Phys. Lett.   {\bf B351}, 99 (1995);
A.D. Linde and A. Mezhlumian, Phys. Rev. D {\bf 52}, 6789 (1995).

\bibitem{HM} S.W. Hawking and I.G. Moss, {  Phys. Lett.}  {\bf 110B},  35
(1982).

\bibitem{Toy} A.D. Linde, Phys. Rev. {\bf D59}, 023503 (1999),   
hep-ph/9807493.

\bibitem{Coleman} S. Coleman,   Phys. Rev. D {\bf 15}, 2929  (1977).

\bibitem{nucl} A.D. Linde,   Nucl. Phys. {\bf B216},  421 (1983);
  Nucl. Phys. {\bf B372}, 421 (1992).

\bibitem{HT} S.W. Hawking and N. Turok, Phys. Lett. {\bf B425}, 25   
(1998),
hep-th/9802030; S.W. Hawking and N. Turok, hep-th/9803156.

\bibitem{ALOpen} A.D.  Linde,   Phys. Rev. D {\bf 58}, 083514 (1998),
  gr-qc/9802038,

\bibitem{BC} R. Bousso and A. Chamblin, gr-qc/9803047.

\bibitem{STYtensor} M. Sasaki, T. Tanaka, and Y. Yakushige, Phys. Rev. D
  {\bf 56}, 616 (1997).

\bibitem{Bell} J. Garc\'{\i}a--Bellido, Phys. Rev. D {\bf 56}, 3225 (1997), 
 astro-ph/9702211.

\bibitem{GMST1} J. Garriga, X. Montes, M. Sasaki and T. Tanaka,
  Nucl. Phys. B {\bf 513}, 343 (1998).

\bibitem{GMST2} J. Garriga, X. Montes, M. Sasaki and T. Tanaka,
astro-ph/9811257.

\bibitem{YST} K. Yamamoto, M.  Sasaki and T. Tanaka, Phys. Rev. D {\bf
    54}, 5031 (1996).

\bibitem{Tomita} K. Tomita, Prog. Theor. Phys. {\bf 68}, 310 (1982).

\bibitem{TStensor} T. Tanaka and M. Sasaki, Prog. Theor. Phys. {\bf 97},
  243 (1997).

\bibitem{Bucher} M. Bucher and J.D. Cohn,
Phys. Rev. D {\bf 55}, 7461 (1997).

\bibitem{COBE} G.F. Smoot et al., Astrophys. J. {\bf 396}, L1 (1992);
  C.L. Bennett et al., Astrophys. J. {\bf 464}, L1 (1996).

 
\end{thebibliography}
\end{document}